\documentclass[pdflatex,sn-mathphys-num]{sn-jnl}

\usepackage{graphicx}%
\usepackage{multirow}%
\usepackage{amsmath,amssymb,amsfonts}%
\usepackage{amsthm}%
\usepackage{mathrsfs}%
\usepackage[title]{appendix}%
\usepackage{xcolor}%
\usepackage{textcomp}%
\usepackage{manyfoot}%
\usepackage{booktabs}%
\usepackage{algorithm}%
\usepackage{algorithmicx}%
\usepackage{algpseudocode}%
\usepackage{listings}%

\usepackage{braket}
\usepackage{mathcomp}
\usepackage{gensymb}
\usepackage{here}
\usepackage{amsmath}
\usepackage{lineno}

\theoremstyle{thmstyleone}%
\theoremstyle{thmstyletwo}%

\theoremstyle{thmstylethree}%

\begin{document}

\title[Article Title]{Experimental evidence for the physical delocalization of individual photons in an interferometer}



\author*[1]{\fnm{Ryuya} \sur{Fukuda}}\email{d254149@hiroshima-u.ac.jp}

\author[1]{\fnm{Masataka} \sur{Iinuma}}\email{iinuma@hiroshima-u.ac.jp}

\author[1]{\fnm{Yuto} \sur{Matsumoto}}\email{m241959@hiroshima-u.ac.jp}

\author[1]{\fnm{Holger F.} \sur{Hofmann}}\email{hofmann@hiroshima-u.ac.jp}

\affil[1]{\orgdiv{Graduate School of Advanced Science and Engineering}, \orgname{Hiroshima University}, \orgaddress{\street{Kagamiyama 1-3-1}, \city{Higashi Hiroshima}, \postcode{739-8530},  \country{Japan}}}


\abstract{It is generally assumed that the detection of a single photon as part of an interference pattern erases all possible which-path information. However, recent insights suggest that weak interactions can provide non-trivial experimental evidence for the physical delocalization of a single particle passing through an interferometer. Here, we present an experimental setup that can quantify the delocalization of individual photons using the rate of polarization flips induced by small rotations of polarization. The results show that photons detected in equal superpositions of the two paths are delocalized when detected in a high probability output port, and ``super-localized'' when detected in a low probability output port. We can thus confirm that delocalization depends on the detection of photons in the output of the interferometer, providing direct experimental evidence for the dependence of physical reality on the context established by a future measurement.
}

\maketitle
\newpage

\section{Introduction}\label{sec1}
Single photon interference is widely used as an illustration of the oddity of the statistical predictions of quantum mechanics \cite{Feynman1965,Tonomura1989,Zeilinger1999,Young}. In the case of a photon passing through a two-path interferometer, the photon is eventually detected in only one of the two output ports, consistent with the idea that a particle can only be in one place at a time. However, the probabilities of detecting the photon in each of the output ports, determined after many photons were individually injected and detected, can only be explained by a wave-like interference of the two paths inside the interferometer. This duality gives rise to an intriguing question. How can wave-like propagation through the interferometer be reconciled with the local detection of photons? The traditional answer is given in terms of a trade-off relation between which-path information of the photon and the visibility of the interference pattern, often associated with the concept of complementarity\cite{Eng1996,Rempe1998,Barbieri2009,Wang2020}. We are told to accept that nothing can be known about the manner in which the photon propagated through the interferometer when the probability of the observed outcomes depends on interference between the paths.

Recently, this traditional position has come under criticism, both on conceptual grounds and for practical reasons. Conceptually, Bell's inequality violations have shown that quantum theory excludes certain models of reality, indicating that some statements about the past of a particle could be derived from the theoretical formalism \cite{Bell,Ono,Frau18,Bru18,Dressel10,Pusey14,Piacen16,Vir24,Gen25}. On the practical side, new setups and new methods of measurement have provided experimental evidence for the non-classical aspects of single particle propagation in quantum mechanics \cite{Vaidman13past,Vaidman13asking,Vaidman14,Griffith16,Lem18,Lem19,Dajka21,Ji23,Hof23threepath}. A noteworthy breakthrough in this direction was the delayed-choice quantum eraser experiment, which showed that it is possible to ``switch'' between wave-like propagation (i.e.\,interference) and particle-like propagation (i.e.\,path information) by using photon entanglement \cite{Hellmuth1987,Scully1991,Eng1999erasure,Walborn2002}. These results confirm that the wave-particle duality cannot be explained by a measurement independent reality. However, they fail to provide any evidence for the behavior of individual photons.
It is indeed difficult to obtain such information, since it must be extracted from the noisy statistics of many detection events. The trade-off relation between interference effects and paths means that any correlations between the two will be obscured by a considerable amount of background noise. Nevertheless it is possible to isolate such correlations in weak measurements, where the post-selection of a specific photon detection event allows us to evaluate conditional averages of complementary quantities \cite{Aharonov1988,Duck1989,Kocsis11,Mas25}. The fact that weak values lie outside the range of values observed in direct measurement has given rise to a long standing controversy regarding the physical meaning of weak measurements and their results \cite{Dre15,Iinuma16,Per22}.  A central problem is the understanding of the measurement process. In that regard, important insights have been gained through Hall's discussion of Ozawa's operator-based definition of uncertainties \cite{Ozawa21,Hall}. Here, weak values appear as optimal estimates for each of the measurement results \cite{Wiseman10,Nii18,Dvo25}. For pure states, Ozawa's theory predicts a measurement uncertainty of zero when the measurement result is given by the weak value associated with a post-selected measurement result. Motivated by this observation, we recently introduced a method that directly quantifies the statistical fluctuations of weak values by using quantum feedback to compensate the effect of a weak interaction between the system and a quantum probe \cite{Hof21}. 
This result was subsequently applied to demonstrate the delocalization of a single neutron between two paths of an interferometer, strongly indicating that weak values describe the delocalization of particles over the paths of an interferometer \cite{Lem22}. However, these results relied on an initial bias between the paths, making it difficult to apply them to the more conventional case of interferences between paths of equal amplitude.
Here, we will address this problem and show that the delocalization of a single particle can be observed experimentally based on a direct observation of polarization flips as proposed in \cite{Hof23}. The experimental evidence reveals a non-trivial dependence of the observed delocalization on the precise phase shift, demonstrating the dependence of physical reality on the specific measurement applied to a quantum system.

The fundamental idea behind the present experiment concerns the effects of a small polarization rotation on vertically (V) polarized input photons. The rate at which such a polarization rotation flips photons into the horizontal (H) polarization depends only on the magnitude of the rotation angle, not on its direction. If we apply opposite polarization rotations in the two paths of an interferometer, photons traveling along either of the two paths will flip at the same rate. 
The only way in which the flip rate can change is if the two rotations act on the same photon, resulting in a combination of rotations that depends on the relative contribution of each rotation to the change of the polarization experienced by a single photon. The polarization flip rate given by the probability of H-polarization P(H) observed in the output of the interferometer thus provides us with direct experimental evidence of the delocalization of individual photons, operationally defined as the proportional effects of local polarization rotations in the different paths on the individual photon. It should be noted that we do not obtain any path information in the measurement, since the flip rate is the same in both paths and we cannot tell which path contributes more and which contributes less. Instead, the rate of polarization flips is determined by the physical distribution of individual photons over the two available paths. 
Delocalization reduces the flip rate due to the cancellation of opposite rotation, while an enhancement of the flip rate suggests an extreme concentration of the photon in one of the paths. The latter may seem counter intuitive at first, since it indicates a distribution of the photon with negative values in one of the paths, allowing for values greater than one in the other. 
Here, the negative sign of the photon density represents a reversal of the rotation direction, causing the polarization rotations to add up. An enhancement of the polarization flip rate above the level observed in each path is experimental evidence of a “super-localization,” whereby individual photons experience a negative fraction of one polarization rotation and a correspondingly enhanced fraction of the other polarization rotation. 
It should be noted that this result is consistent with the observation of negative weak values in the neutron interference experiment \cite{Lem22}, where it was found that a particle could have a negative presence in the corresponding path, and a presence greater than one in the other. In the present experiment, we show that this super-localization has a direct experimentally observable effect in the rate of spin flips induced by the opposite polarization rotations in the two paths.

The results we present in the following show that the degree of delocalization of each photon detected in the output of the interferometer depends on the output port at which it is detected. For photons detected in the high probability output where interference between the paths is constructive, the flip rate is suppressed indicating a delocalization over the two paths. 
For probabilities close to 100\%, this delocalization corresponds to a perfect splitting of the photon, with exactly half of the photon passing along each of the two paths. As interference effects weaken, the delocalization of the photon weakens as well, with each photon splitting into different fractions experiencing the polarization rotations in the two paths. 
Oppositely, photons detected in the low probability output where interference between the paths is destructive, our results show an enhancement of the polarization flip rate corresponding to super-localization in one of the paths. This effect reaches a maximum limited only by the visibility of interference. We point out that super-localization is necessary to compensate for the delocalization observed in constructive interferences, since the total rate of polarization flips of the two output ports must always be consistent with the localization of photons in one of the paths observed in a which-path measurement.

Our experiment demonstrates that the past of a quantum particle depends on the future measurements by which the particle is ultimately detected. In a somewhat ironic twist, this result could confirm a claim often made in the early days of quantum mechanics, where it was suggested that the outcomes of measurements depend on the participation of the observer. 100 years later, we may finally have found objective evidence that can explain the meaning of this cryptic claim. The reason for this breakthrough are the new experimental possibilities that allow us to extract maximal information from the observation of weak effects. Any conventional detection of particles in the paths is a massive intervention that destroys all information about possible interference effects, leaving only philosophical speculations regarding the relation between two separate experimental scenarios. On the other hand, weak interactions create a quantum memory of the causes of future interference effects, revealing a correlation between the distribution over the paths and the phase relations that determine the output port in the interference experiment. This correlation proves that interference effects require a physical delocalization of each particle, as classical wave theory would suggest. Whether a particle is localized or delocalized in any given set of paths is not determined by the initial state alone, but depends equally on the measurement that completes the observable effects of each individual particle. Our results thus indicate that a better understanding of measurement processes is needed to properly explain the physical meaning of quantum states in the wider contexts of new experimental possibilities.

\section{Observable effects of photon delocalization}\label{sec2}

\begin{figure}[h]
\centering
\includegraphics[width=1\linewidth]{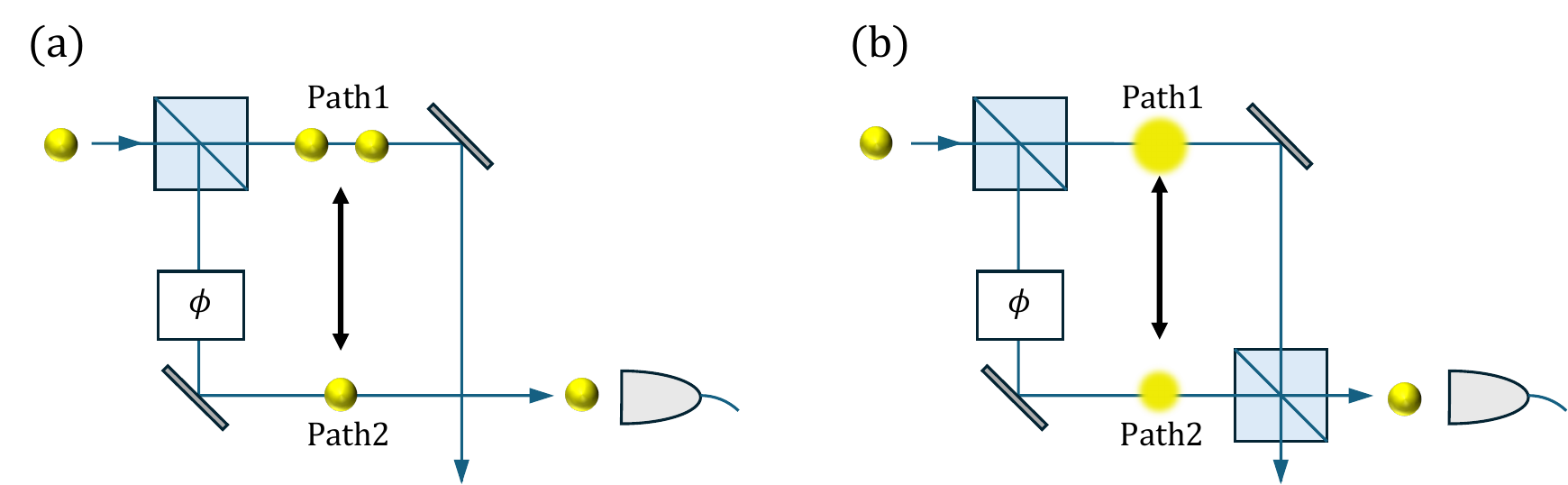}
\caption{Illustration of the measurement dependence of photon delocalization. (a) shows localized photons detected in path 1 or path 2. At the initial beam splitter, the path of each photon is selected randomly. (b) shows a delocalized photon detected after interference at a second beam splitter. In this case, it is conceivable that the photon is physically delocalized, with a larger part of the photon in one path and a smaller part in the other. The photon physically separates into two quantities that propagate along the two different paths.}\label{Fluctuation_Set}
\end{figure}

The quantum formalism provides no clear explanation of the concept of superposition. When a single photon input passes through a beam splitter, the amplitudes of the wave functions can be interpreted as detection probabilities for the two paths behind the beam splitter, as shown in Fig.\,\ref{Fluctuation_Set}(a). However, both amplitudes will be necessary to determine the output probabilities when a second beam splitter interferes the two path, as shown in Fig.\,\ref{Fluctuation_Set}(b). Interference effects thus suggest that the photon can be physically delocalized inside the interferometer, with part of the photon in one path and another part in the other. 

\begin{figure}[H]
\centering
\includegraphics[width=0.9\linewidth]{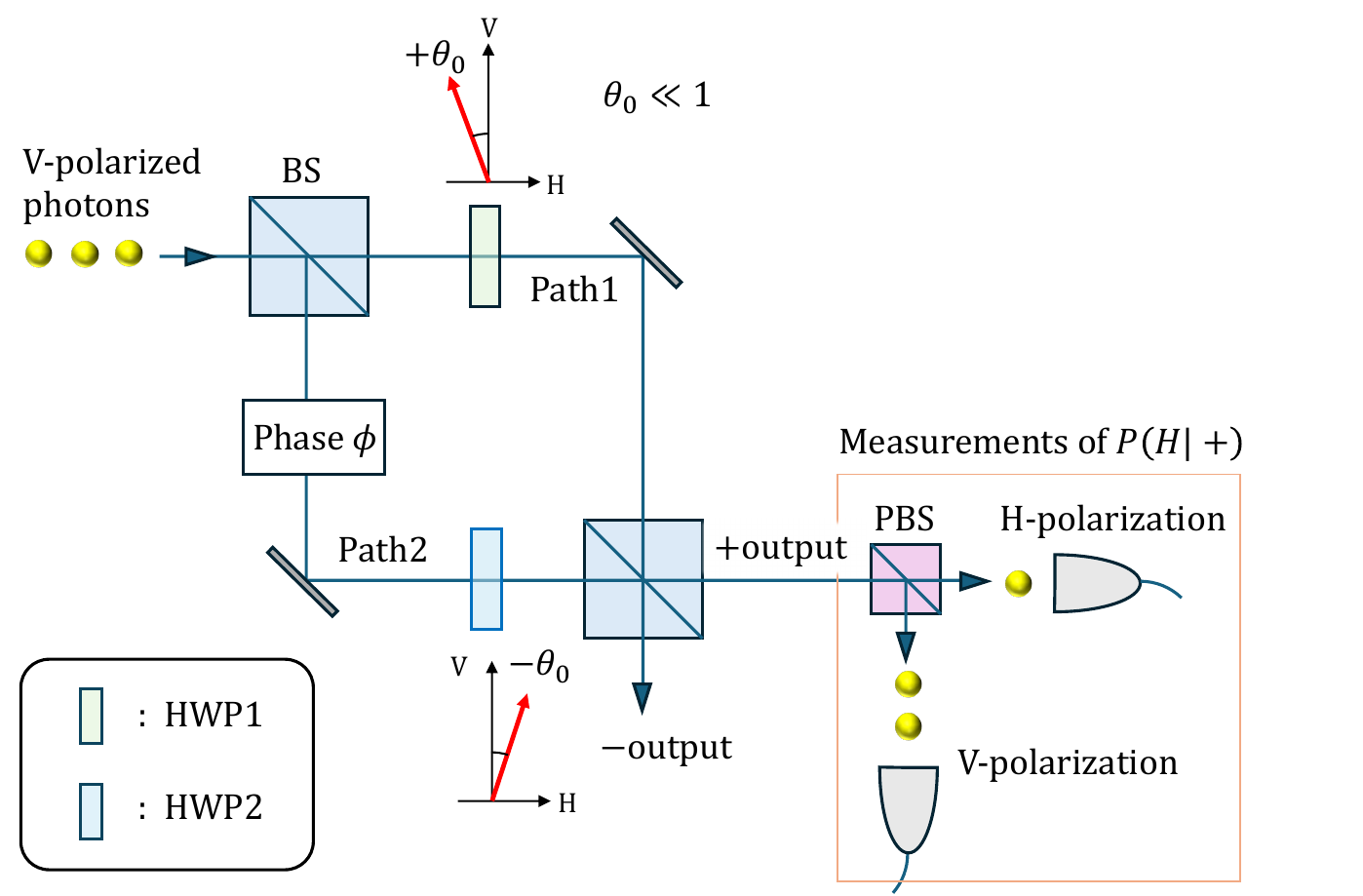}
\caption{Method for observing photon delocalization. Vertically polarized photons are injected into a two-path interferometer. We apply local operations to the polarization using two Half-Wave-Plates(HWPs), HWP1 and HWP2, placed into the two paths of the interferometer. HWP1 rotates the polarization by a small angle $\theta_0(\ll 1)$, and HWP2 rotates the polarization in the opposite direction by the same angle $-\theta_0 (\ll 1)$. Since the probability of a polarization flip from V-polarization to H-polarization is proportional to the square of the rotation angle, localized photons all flip with the same probability, $P(H|1)=P(H|2)$. When interference is observed, the polarization flip probabilities $P(H|\pm)$ observed in the output ports change, where lower flip probabilities indicate that the local rotations can cancel each other. The flip probabilities $P(H|\pm)$ thus provide direct evidence for the delocalization of photons inside the interferometer.}  
\label{Ma_Set}
\end{figure}

Here, we show that it is possible to obtain experimental evidence for this delocalization effect by evaluating the effects of local polarization rotations on the polarization observed in the output of the interferometer. For this purpose, we consider the interaction between the paths and the polarization quantitatively by assigning values of $+1$ to path 1 and $-1$ to path 2. The path quantity is then given by 
\begin{equation}
\label{eq:A_hat}
    \hat{A} = \ket{1}\!\bra{1} - \ket{2}\!\bra{2}.
\end{equation}
The eigenvalues of the quantity $\hat{A}$ represent the presence of the photon in path 1 or path 2. If the initial state is the eigenstate $\ket{1}$, $\hat{A}$ returns the eigenvalue $+1$, representing a photon localized in path 1. If the eigenstate is $\ket{2}$, $\hat{A}$ gives the eigenvalue $-1$, representing a photon localized in path 2. The eigenvalues of $\hat{A}$ thus represent the localization of photons in the paths. Oppositely, any observation of values other than $\pm 1$ would be an indication that the photons are delocalized in some form.

We would like to emphasize that we intend to identify the magnitudes of individual values of $\hat{A}$, not merely the average values. This is the reason why we cannot simply implement a weak measurement where the weak value of $\hat{A}$ is obtained from the average of weak measurement outcomes \cite{Dressel10}. Instead, we implement a post-selected measurement of the squared value of $\hat{A}$, where the positive average provides us with reliable evidence for the magnitude of $\hat{A}$ conditioned by the post-selected measurement outcome \cite{Hof23}. 
We consider an input state represented by an equal superposition of the two paths, $\ket{\psi} = \frac{1}{\sqrt{2}}(\ket{1}+\ket{2})$. If the photon is detected in one of the output ports of the interferometer, the photon cannot be detected in either of the paths and no eigenvalues can be assigned to the quantity $\hat{A}$. However, we can still obtain information about the value of $\hat{A}$ by transferring path information to the polarization of the photon. Starting with vertically polarized photons, we rotate the linear polarization by an angle of $\theta_0 \hat{A}$, that is, we apply a local rotation of $+\theta_0$ in path 1 and a local rotation of $-\theta_0$ in path 2. Such polarization rotations can be realized by Half-Wave-Plates (HWPs) in the paths as indicated in Fig.\,\ref{Ma_Set}. If the local angle $\theta_0$ is sufficiently small $(\ll 1)$, these rotations have very little effect on the interference observed in the output ports. The value of $\hat{A}$ is now encoded in the rotation angle of the photons, and the effects of this rotation by an angle of $\theta_0 \hat{A}$ can be observed by detecting the polarization of the photons in the output of the interferometer.

Since our intention is to distinguish between delocalized and localized photons, we are mostly interested in the magnitude of $\hat{A}$ given by its square. This magnitude can be observed in the probability $P(H|\pm)$ that the polarization flips from V-polarization to H-polarization. For a small rotation angle $\theta$, the probability of such a polarization flip is approximately given by $\theta^2$. Since the rotation angle is related to the path value by $\theta=\theta_0 \hat{A}$, we can obtain the value of $A^{2}(\pm)$ by 
\begin{equation}
\label{eq:A_square}
    A^2(\pm) = \frac{P(H|\pm)}{\theta_0 ^2}.
\end{equation}
If the photons are localized in either path, the flip probability $P(H|\pm)$ must be equal to $\theta_0^2$. For experimentally observed values of $P(H|\pm) < \theta_0^2$, we find that $A^{2}(\pm)<1$. The rotation angle is smaller than the local rotation angles, indicating a physical distribution of the photon over both paths, so that the opposite rotations partially cancel. In particular, $A^{2}(\pm) \approx 0$ corresponds to an equal distribution where the photon is half in one path and half in the other. A suppression of polarization flips from V-polarization to H-polarization is a reliable indicator of delocalization, allowing us to identify the delocalization associated with the two output ports of the interferometer. 

The physical effect of delocalization is a reduction of the polarization rotation and the associated flip probability $P(H|\pm)$. Oppositely, an increase of the flip probability $P(H|\pm)$ indicates an increase of the squared rotation angle $\theta^2$ beyond the local limit of $\theta_0^2$. This increase suggests that one of the two rotation angles must have changed its sign. Since the interaction between path and polarization indicates that the polarization angle is proportional to the fraction of the particle present in the path, a change of rotation direction corresponds to a negative fraction of the photon in one of the paths. This negative fraction will be compensated by a corresponding fraction greater than one in the other path, resulting in an intensity difference between the paths that is greater than the photon number of one. When viewed as a single particle, the photon is effectively "super-localized" in one of the paths, with a negative fractional presence in one path allowing for a much greater fluctuation in $\hat{A}$ than the limit for localized photons. In all cases, the magnitude of the quantity $\hat{A}$ is directly observed in the probability of the polarization flips, where values of $P(H|\pm) < \theta_0^2$ indicate delocalization and values of $P(H|\pm) > \theta_0^2$ indicate super-localization.

\section{Experimental Setup}\label{sec3}

\begin{figure}[H]
\centering
\includegraphics[width=1\linewidth]{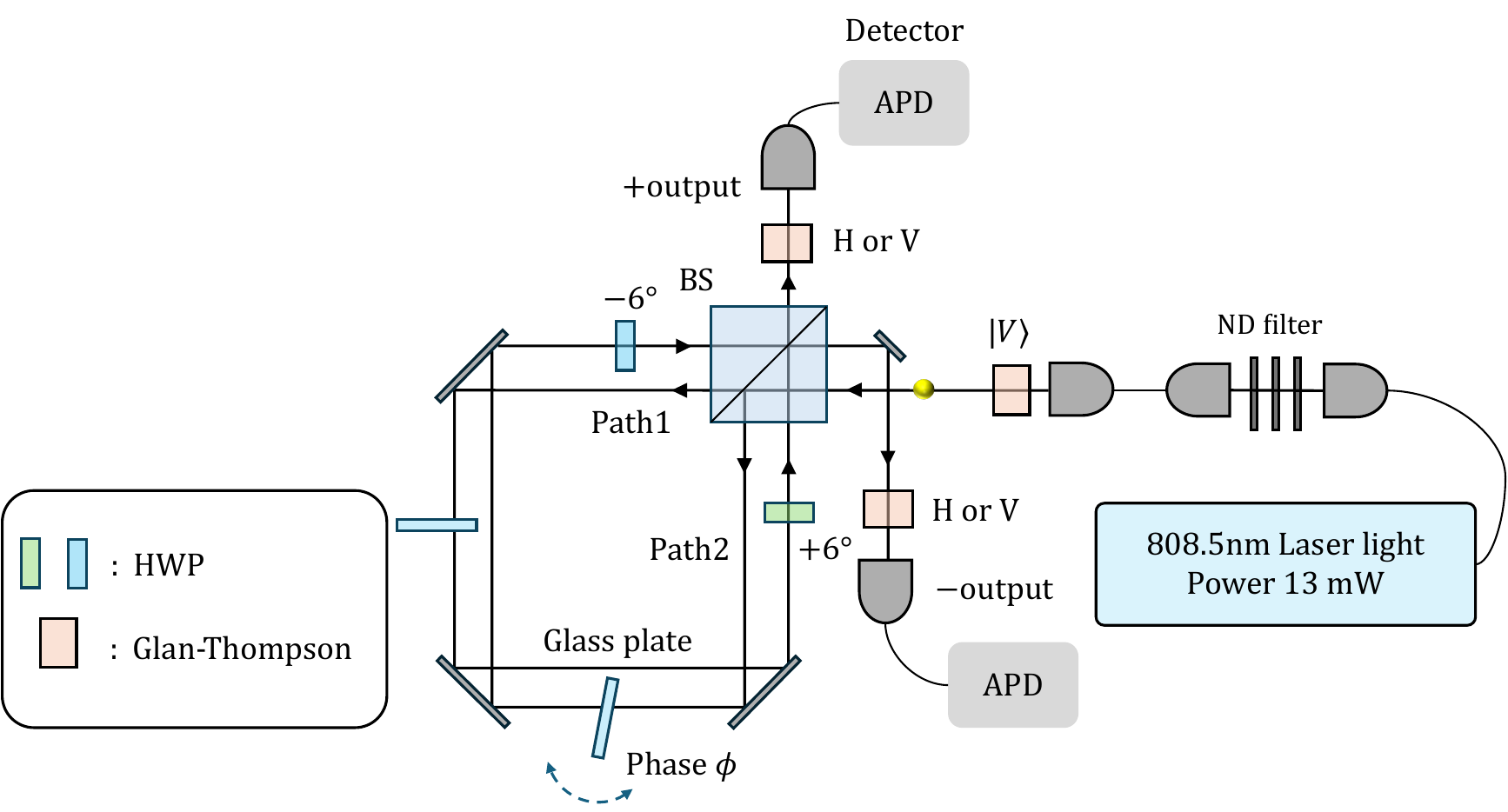}
\caption{Illustration of the experimental setup with a Sagnag-like interferometer. The vertically polarized input photons were emitted by a laser, which is weakened using an ND filter to obtain a photon rate of 110000/s on average.
A beam splitter(BS) with a 50:50 split was used for preparation of the superposition state. The relative phase $\phi$ was controlled by tilting either of two glass plates placed on two paths in the interferometer. The output photons were counted by two Avalanche photo detectors(APD) for 100s at each phase, where Glan-Thompson Polarizers (GT) were used to distinguish horizontal and vertical polarization. }
\label{Sa_Set}

\end{figure}
Our experimental setup is shown in  Fig.\,\ref{Sa_Set}. Its central part is a Sagnac-like interferometer that ensures the stability of the relative phase between the two paths. 
As photon source, we used a laser diode with a wavelength of $808.5 \mbox{nm}$. The intensity was reduced by ND filters to a photon rate of 110000/s on average. The input photons enter and exit the interferometer through a 50:50 beam splitter (BS).  Phase shifts between the paths were controlled using glass plates inserted into only one path each. The polarization of the output photons were discriminated with two Glan-Thompson Polarizers (GT) placed on each output just before a fiber coupler connected to an avalanche photo diode (APD) detector through an optical fiber.   
The dark counts of both APDs were evaluated by blocking the light from the photon source, allowing us to subtract the dark count background from the total counts to achieve accurate count rates at low detection probabilities (details in Appendix B). 

We found that the main source of experimental imperfections was the beam splitter through which the photons enter and exit the interferometer. The precise ratio of reflection and transmission at the beam splitter was 1.121, which corresponds to a reflectivity of 52.85 \%. We note that this imbalance between transmission and reflection has a negligible effect on the visibility of the interference between the path. More problematic is the polarization dependence of this imbalance. To avoid errors in the measurement of delocalization, we blocked one path to characterize the rotated polarization of that path in the output. We could then confirm that the polarizations were rotated in opposite directions with a slight adjustment of the two GTs (details in Appendix A1.2). 

As explained in the previous section, photon polarization serves as a probe of the spatial delocalization of each photon between the paths. Specifically, the probability that a V-polarized input photon flips to H-polarization determines the value of $A^2(\pm)$ and therefore the delocalization of photons exiting the interferometer at the $+$ and the $-$ ports, respectively. Data was taken while varying the phase difference $\phi$ in 32 steps for a full range of $180\degree$, each step being about $5.625\degree$ in length. We switched the direction of the detected polarization between H and V, obtaining count rates for each polarization direction using the same time window 100s for each setting. This data was then used to evaluate the flip probabilities $P(H|+)$ and $P(H|-)$ when the photon is detected in a specific output port.

\section{Results}\label{sec4}
\begin{figure}[tbhp]
    \centering
    \includegraphics[width=0.7\linewidth]{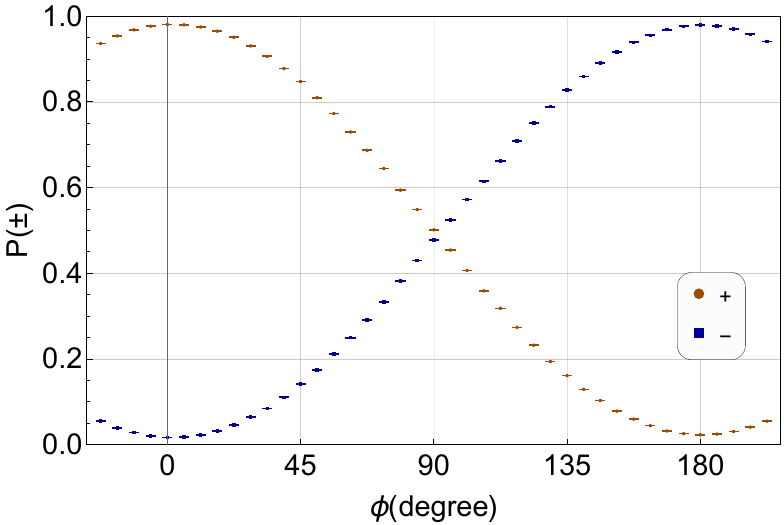}
    \caption{Probabilities $P(+)$ and $P(-)$ of detecting the photons in the respective output ports of the interferometer at different phases $\phi$. The phase was changed $-22.5\degree$ to $202.5\degree$ in steps of $5.625\degree$. The visibilities obtained from the data are 0.9575 for the $+$ output and 0.9629 for the $-$ output. These visibilities include the decoherence effects induced by the local polarization rotations in the paths.}
    \label{PM_result}
\end{figure}

We first characterized the interference fringes of the Sagnac-like interferometer by combining the counts of the H and V polarized photons at the output ports to obtain the total output probabilities $P(+)$ and $P(-)$. The interference fringes obtained in this measurement are shown in Fig.\,\ref{PM_result}. Note that the polarization rotations in the paths have already been implemented. The experimentally observed visibility thus includes all decoherence effects caused by the interactions between polarizations and paths. The interference visibilities were evaluated to be 0.9575 for the $+$ output and 0.9629 for the $-$ output. The slight difference can be explained by the effects of imperfections of the beam splitter used in the interferometer, which have less impact on the $P(+)$ result at $\phi=0\degree$ than on the $P(-)$ result at $\phi=180\degree$. 

\begin{figure}[H]
    \centering
    \includegraphics[width=1\linewidth]{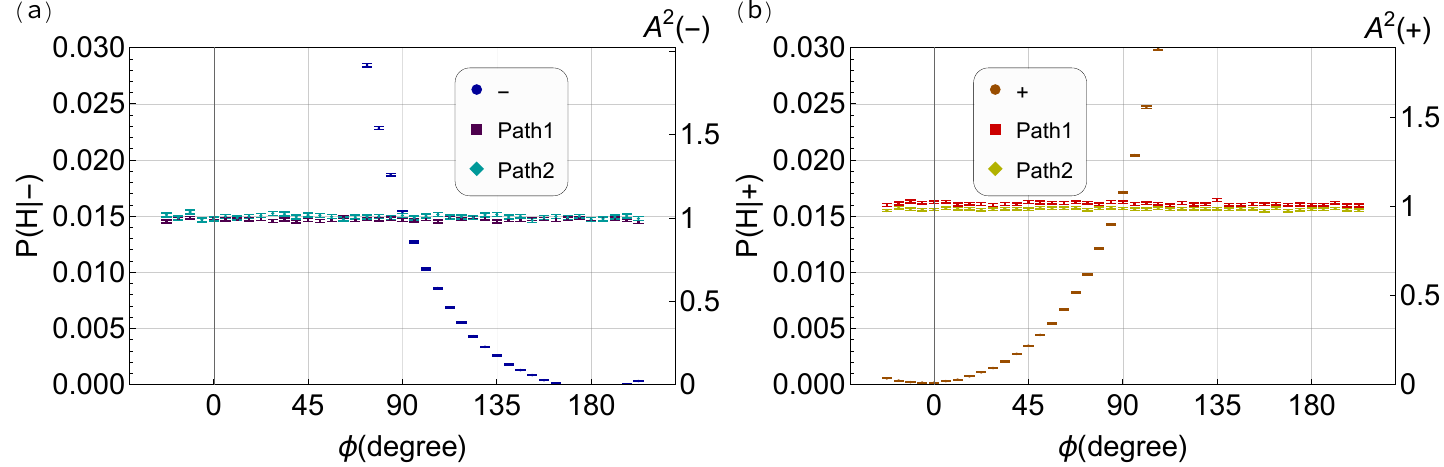}
    \caption{Delocalization of photons observed in output ports preferred by constructive interference. Graph (a) shows the phase dependence of $P(H|-)$, where constructive interference is observed for $\phi>90\degree$. Graph (b) shows the phase dependence of $P(H|+)$, where constructive interference is observed for $\phi<90\degree$. The solid circles represent the measurement results when the two paths interfered at the output. Solid squares and solid diamonds represent the data obtained when one of the paths was blocked. This data represents a value of $A^2(\pm)=1$, characteristic of localized photons. The axes on the right side of the graphs give the corresponding values of $A^2(\pm)$ based on this comparison. Delocalization is directly observed whenever constructive interference favors the output port in which the photon was detected.}
    \label{PH_zoom}
\end{figure}

\begin{figure}[H]
    \centering
    \includegraphics[width=0.98\linewidth]{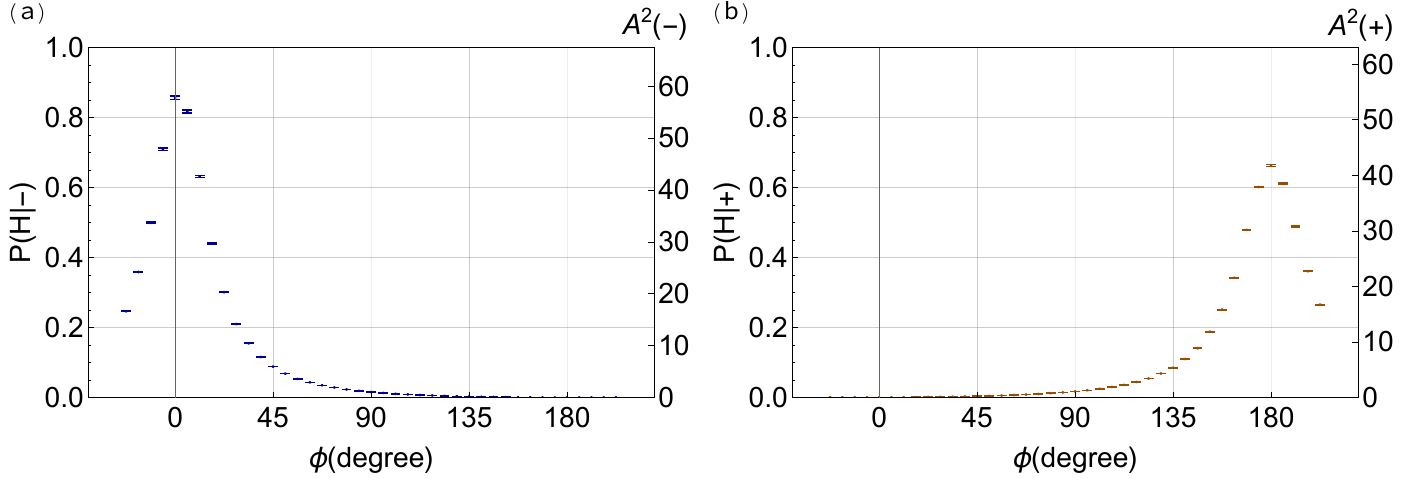}
    \caption{Super-localization of photons observed in output ports where the output probability is suppressed by destructive interference. Graph (a) shows the phase dependence of $P(H|-)$, where destructive interference is observed for $\phi<90\degree$. Graph (b) shows the phase dependence of $P(H|+)$, where destructive interference is observed for $\phi>90\degree$. The axes on the right side of the graphs give the corresponding values of $A^2(\pm)$. The flip probability $P(H|\pm)$ reaches its maximal value where the output probabilties $P(\pm)$ are minimal. $P(H|-)$ achieves a maximal value of $0.857\pm0.005$ at $\phi=0 \degree$ and $P(H|+)$ achieves a maximal value of $0.663\pm0.002$ at $\phi=180\degree$. These maximal values are consistent with the visibilities of 0.9629 for the $-$ output and of 0.9575 for the $+$ output. The difference between the two maximum values can thus be explainable by the slight difference of the visibility of destructive interference in the two output ports. In terms of the $A^{2}(\pm)$, the maximal observed super-localizations are $A^2(-)=57.80\pm0.34$ at $\phi=0\degree$ and $A^2(+)=41.78\pm0.17$ at $\phi=180\degree$, roughly corresponding to a four-fold enhancement of the polarization rotation in one path, and a reversal and three-fold enhancement of the rotation in the other path.}
    \label{PH_all}
\end{figure}

Fig.\,\ref{PH_zoom} and Fig.\,\ref{PH_all} show the experimental results for the flip probabilities $P(H|\pm)$ and the corresponding square of the path value $A^2(\pm)$. In Fig.\,\ref{PH_zoom}, the polarization flip rates obtained for photons localized in one of the two paths are compared with the flip rates observed for photons observed after the two paths interfered. The data for the individual paths was taken by blocking the other path, while leaving all other settings unchanged. The results taken for photons localized in path 1 and in path 2 show that localized photons can only induce a polarization flip probability of about 0.015. The values obtained by averaging over the results obtained for different phase shifts are $0.01482\pm0.00003$ for output port $-$ and $0.01587\pm0.00003$ for output port $+$ (details in Appendix A1.2). When both paths are open and interference occurs, the situation changes significantly, as shown by the solid circles in Fig.\,\ref{PH_zoom}. As shown in Fig.\,\ref{PH_zoom} (a), a flip rate of $P(H|-)<0.015$ ($A^2(-)<1$) is observed for $\phi > 90\degree$, indicating that the photons are physically delocalized between the paths whenever constructive interference favors the $-$ output.  The flip probability $P(H|-)$ drops to nearly zero around $\phi=180\degree$, indicating that the polarization rotations in the two path cancel each other. This cancellation can only be achieved when each photon interacts equally with both HWP1 and HWP2. Quantitatively, the photons must be physically distributed over the two path, with half of the photon in one path, and half in the other. Likewise, as shown in Fig.\,\ref{PH_zoom} (b), a flip rate of $P(H|+)<0.015$ ($A^2(+)<1$) is observed for $\phi < 90\degree$, indicating delocalization of the photons whenever constructive interference favors the $+$ output.  The flip probability $P(H|+)$ drops to zero around $\phi=0\degree$, indicating maximal delocalization with half a photon in one path and half in the other. 

Fig.\,\ref{PH_zoom} shows that photons will be delocalized when they exit the interferometer in a port favored by constructive interference. Delocalization is thus associated with the most likely output detection results. Oppositely, photons in the output port where the output probability is suppressed by destructive interferences have flip probablities of $P(H|\pm)>0.015$, corresponding to values of $A^2(\pm)>1$. As discussed before, this increase of the flip probabilities above the value for localized photons indicates a form of ``super-localization,'' where a negative fraction of the particle in one path is needed to explain the effective reversal of the direction of the polarization rotation in that path, while single particle normalization requires that a fraction of the particle that is greater than one experiences the rotation in the other path. It may be worth noting that this result is consistent with the weak values of $\hat{A}$ that have been observed with an input state biased in favor of one of the paths \cite{Lem22}. Fig.\,\ref{PH_zoom} shows that super-localization is observed in one output port when delocalization is observed in the other. This is consistent with the idea that $A^{2}(\pm)$ is proportional to the probability of H-polarized photons. Since the interference effect at the final beam splitter cannot change the total number of H-polarized photons, the statistical average of $A^2$ over all outcomes must be independent of the final measurement, whether it be a path measurement or an interference measurement. Since both individual values and the average are equal to one in the path measurements, the average of the interference measurement must likewise be one. Therefore, 
\begin{equation}
\label{eq:constant}
  A^2(+)P(+) + A^2(-)P(-) =1
\end{equation}
for all phases $\phi$. This relation has an interesting consequence for the upper limits of super-localization. Close to zero degrees, $A^2(+)$ is close to zero, but the probability $P(-)$ is also close to zero. It follows that, for $\phi=0$, $A^2(-) \approx 1/P(-)$. The maximal observable super-localization thus depends on the visibility of interference. 

Fig.\,\ref{PH_all} shows the experimental results for flip probabilities $P(H|\pm)>0.015$  ($A^2(\pm)>1$), indicating super-localization. As shown in Fig.\,\ref{PH_all} (a), $P(H|-)$ peaks at a maximal value of $0.857\pm0.005$ at $\phi=0\degree$, corresponding to a value of $A^2(-)=57.80\pm0.34$. Eq.(\ref{eq:constant}) associates this value of $A^2(-)$ with a probability of $P(-)=0.0173$, corresponding to a visibility of 0.965. The maximal value of $A^2(-)$ determined from the flip probabilities $P(H|-)$ is therefore consistent with the visibility of 0.9629 experimentally determined from the output probabilities $P(\pm)$ at $\phi=0\degree$. Fig.\,\ref{PH_all} (b) shows the results for $P(H|+)$, where a maximal value of $0.663\pm0.002$ is obtained at $\phi=180\degree$. The corresponding value of $A^2(+)$ is $41.78\pm0.17$. The output probability estimated from the inverse $1/A^2(+)$ is 0.024, equivalent to a visibility of 0.952. The result is consistent with the visibility of 0.9575 experimentally determined from the output probabilities $P(\pm)$ at $\phi=180\degree$. The difference between the maximal values of super-localization at $\phi=0\degree$ and at $\phi=180\degree$ can thus be explained by the different visibilities observed at these phase shifts.   

The results shown in Fig.\,\ref{PH_all} show that the maximal observable super-localization in our experiment is around $A^2(\pm)=50$, associated with a visibility of about 0.96. This translates to absolute values of $\hat{A}$ of around seven, where the polarization rotation in one path is enhanced four-fold and the negative rotation in the other path is enhanced three-fold. This is the amount of super-localization in one of the paths that is needed to explain how the experimentally observable effect of the local polarization rotations can be enhanced by a factor of around 50.

\section{Discussion}\label{sec5}

Our results show how individual photons detected in the outputs of an interferometer are physically delocalized as they propagate through the interferometer, where the concept of delocalization is explained in terms of the empirical evidence provided by local interactions within the paths. By ensuring that the polarization flip rate is the same for all photons localized in only one of the paths, we identify an experimentally observable phenomenon directly associated with the delocalization of the particles between the two paths. As shown in Fig.\,\ref{PH_zoom}, this phenomenon is consistent with the fact that particles are localized in one of the paths whenever they are detected in that path. However, the situation changes when the particles are detected in the output ports of the interferometer instead. The majority of photons detected in the output port favored by constructive interference experiences a proportional combination of both local effects, indicating a corresponding delocalization of the photons within the interferometer. The stronger the interference effect, the more equal the distribution of each photon between the paths. This effect is consistent with the expectation that interference works best when the intensities of both beams are equal. Photons detected in constructive interference thus behave very much like classical wave intensities. However, their detection in the output port is still a particle-like all-or-nothing event, as demanded by the statistical interpretation of quantum mechanics.

The observed values for the conditional delocalization of a single photon between the paths demonstrate that the relation between statistics and physical quantities cannot be reduced to the naive concept of complementarity suggested in early experiments such as \cite{Rempe1998}. Instead, there is a causal relation between the quantum fluctuations of $\hat{A}$ and the output statistics of the interferometer. Delocalization causes the photon to exit in the port favored by constructive interference and super-localization causes the rare events of a photon exiting in the other port. Eq.(\ref{eq:constant}) indicates that these fluctuations originate from the same fluctuation of $\hat{A}$ that is also observed in which-path measurements where the squared value of $\hat{A}$ is one in each of the two paths. Fig.\,\ref{PH_zoom} shows how the same polarization rotations can indicate either localization or delocalization and super-localization, depending only on the measurement performed after the polarization rotations have been applied. This result illustrates how reality can be measurement dependent. Quantum fluctuations have no intrinsic reality because their specific values depend on their ultimate macroscopic effect. Without such a macroscopic effect in mind, it is not possible to visualize the physics of the system. 

This does not mean that the effects of delocalization and super-localization are less real than the localization observed in the direct detection of a particle. In particular, super-localization describes the necessary cause of the unlikely transmission of individual photons to the low probability output that is suppressed by destructive quantum interference. Fig.\,\ref{PH_all} shows the dependence of super-localization on the phase setting that determines the amount of destructive interference. Maximal values of $A^2$ are obtained when destructive interference reduces the output probability to its minimal value. Here, the small polarization rotations used to evaluate super-localization have a noticeable effect, limiting the observable super-localization by introducing a small amount of decoherence. Note that this does not mean that super-localization depends on the interaction used to observe it. In principle, the rotation angle could always be reduced further, demonstrating the inverse relation between super-localization and detection probability for arbitrarily small probabilities. 

Super-localization is a fundamental quantum phenomenon associated with the post-selection of destructive quantum interference. We find that this effect is closely related to post-selected metrology, where phase sensitivity can be enhanced beyond the limit set by the quantum uncertainty of the initial state \cite{Arv20,Arv24,Zhu25}. Indeed, the theory of super-localization in appendix C is very similar to the theory of enhanced phase sensitivity presented in \cite{Arv20,Arv24}. We are currently working on an experimental realization of a metrology scenario where super-localization is applied to enhance the phase sensitivity of a single photon by a factor corresponding to the enhancement of $A^2$ observed in Fig.\,\ref{PH_all}, demonstrating the usefulness of super-localization as a useful resource in quantum metrology. 

To summarize our results, we have shown that the physical delocalization of individual particles inside an interferometer can be determined experimentally by using sufficiently weak interactions. The results show that particles are only localized when the final measurement is not sensitive to interference effects. These findings have serious implications for the way in which we think about quantum superpositions. In the absence of empirical evidence to the contrary, it is indeed tempting to claim that each component of a quantum state represents the outcome of a measurement, whether the measurement is performed or not. It is therefore important to recognize that our results demonstrate the fallacy of this counterfactual assumption. In the formalism of quantum mechanics, the role of state vectors and their superpositions is ambiguous. Only an actual experiment can identify the physical meaning of such concepts. Misunderstandings may have arisen from a rather hasty identification of physical properties with the eigenstates of operators that represent them in the formalism. As our previous work has shown\cite{Hof21,Lem22,Hof23}, the experiment presented here is consistent with the standard formalism of quantum mechanics, and the definition of delocalization used here can be formalized in terms of the appropriate operator ordering (see details in Appendix C). In the present work, we have highlighted the experimental realization because a purely mathematical definition of fundamental concepts would seem arbitrary and carry very little practical meaning. The problem of the conventional approach to quantum mechanics is that its abstractions hide the wide range of possibilities by which we can explore nature. The present work solves this problem for the case of individual particles in an interferometer, paving the way towards a better practical understanding of all quantum phenomena.

\vspace{1cm} 

\noindent
\textbf{Acknowledgements}
H.F.H. was supported by ERATO, Japan Science and Technology Agency (JPMJER2402);
R. F.  was supported by JST SPRING, Grant Number JPMJSP2132;\\

\newpage






\appendix
\setcounter{section}{1}
\setcounter{subsection}{0}
\setcounter{figure}{0}
\setcounter{table}{0}
\setcounter{equation}{0}

\renewcommand{\thesection}{A\arabic{section}}
\renewcommand{\thesubsection}{A\arabic{section}.\arabic{subsection}}

\renewcommand{\thefigure}{A\arabic{figure}}
\renewcommand{\thetable}{A\arabic{table}}
\renewcommand{\theequation}{A\arabic{equation}}

\section*{Appendix A. Reduction of systematic effects}\label{secA1}

\subsection{Elliptical polarization components}

We found that imperfections of the BS used in the experiment introduced a small amount of elliptical polarization, corresponding to an additional phase shift $\phi_{HV}$ between the H and the V components. If this phase shift is the same for both paths, it causes a systematic error in the $P(H|\pm)$ measurements. To cancel out the elliptical components in the transmitted and reflected output paths, two HWPs placed in the interferometer were individually tilted towards the optical axis. The magnitude of the elliptical component $\langle S_{RL} \rangle$ was directly measured at the $+$ and the $-$ ports for path 1 and path 2 by blocking the respective other path in the interferometer. The adjustment of each HWP and the measurement of the elliptical component were iterated to reduce the sum value of the two elliptical polarization components. The measurement results for $\langle S_{RL} \rangle$ obtained with the configuration used in the experiment are summarized in Supplementary Table \ref{tab:elliptical}. 
\begin{table}[h]
\renewcommand{\tablename}{Supplementary Table}
    \centering
    \caption{Summary of measurement results \\of elliptical components $\langle S_{RL} \rangle$}
    \label{tab:elliptical}
    \begin{tabular}{ccccc}
    \hline
      &  path 1  &  path 2  &  summation  & offset angle \\ 
      &          &          &  $\Delta \langle S_{RL} \rangle $ &  $\Delta \phi$ \\ \hline
    $+$ port & $0.0759\pm0.0007$ & $-0.0604\pm0.0006$ & $0.0155\pm0.0009$ & $3.52\pm0.2\degree$\\ 
    $-$ port & $0.0695\pm0.0007$ & $-0.0698\pm0.0008$ & $-0.0003\pm0.0011$ & $-0.07\pm0.2\degree$\\ \hline
    \end{tabular}
\end{table}

The remaining sum value of the elliptical components $\Delta \langle S_{RL} \rangle$ of the transmitted and the reflected paths results in a phase offset in the horizontal axis of Fig.\,5. The phase offset $\Delta \phi$ in Supplementary Table \ref{tab:elliptical} was evaluated using the relation $\Delta \langle S_{RL} \rangle / 2\theta_{0}$, where the elliptical component can be expressed as $\sin 2\theta_{0} \sin \phi_{HV} \approx 2\theta_{0} \phi_{HV}$. 
The overall effect of the systematic error is sufficiently small, but it is noticeable that the outcomes of the $P(H|-)$ measurement have a lower systematic error than the outcomes of the $P(H|+)$ measurement. 

The difference between the systematic errors is particularly noticeable in Fig.\,5, where the phase offset $\Delta \phi$ shifts the crossing points of the single-path measurements and the interference measurements. For $P(H|-)$, the crossing point is almost precisely at $\phi=90\degree$, whereas for $P(H|+)$ it is not. The mismatch for $P(H|+)$ shown in Fig.\,5 (b) is consistent with the offset angle of $\Delta \phi=3.52\degree$, which roughly corresponds to 2/3 of one step of the data points shown in Fig.\,5 (b).

\subsection{Additional polarization rotation by the final BS}

Another systematic effect originates from an additional polarization rotation caused by an unintended polarization dependence of transmission and reflection at the BS. This additional polarization rotation introduces a difference in the magnitude of the rotation observed in the two output ports that is not related to the delocalization of the photon. To avoid this systematic effect, the orientations of the GTs in the output ports were slightly adjusted so that the values of $P(H|1)$ and $P(H|2)$ observed when the other path was blocked had the same values in both output ports. This adjustment is necessary to satisfy the condition that the magnitude of the polarization rotation is determined by the opposite polarization rotations in the paths only, with no systematic dependence on the output path. To evaluate the compensation angle, we measured the dependence of $P(H|1)$ and $P(H|2)$ on the angle of the GT $\theta_{GT}$ by blocking either of the two paths as shown in Supplementary Figure \ref{GTalign_result} and analyzed the results by fitting the parameters. 
\begin{figure}[h]
\renewcommand{\figurename}{Supplementary Figure}

    \centering
    \includegraphics[width=1\linewidth]{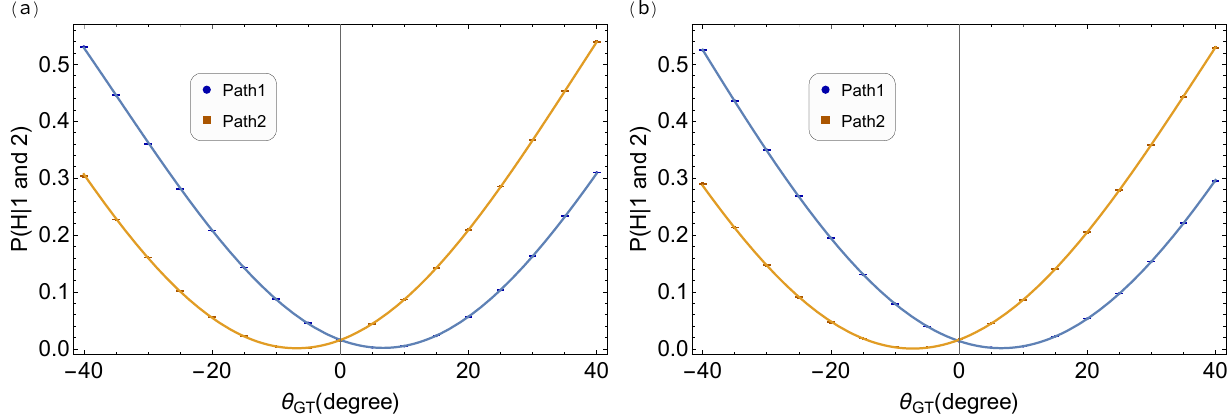}
    \caption{Dependence of $P(H|1)$ and $P(H|2)$ on the $\theta_{GT}$. Graph (a) shows the results measured at the $+$ output port and graph (b) shows the results at the $-$ output port. The evaluation of the two curves by parameter fitting shows that the condition of $P(H|1)=P(H|2)$ is satisfied for a compensation angle of $0.0407\degree$ for the $+$ output port and an angle of $-0.2739\degree$ for the $-$ output port. }
    \label{GTalign_result}
\end{figure}
Graph (a) in Supplementary Figure \ref{GTalign_result} shows the results measured at the $+$ output port and graph (b) shows the results at the $-$ output port. The fitting function is 
\begin{equation}
P(H|1 \; \text{or} \; 2)=A \cos(n(\theta_{GT}-\theta_{GT0}))+A_{0},  
\end{equation}
where the $A$, $n$, $\theta_{GT0}$, $A_{0}$ are fitting parameters. The correct compensation angles should satisfy the condition of $P(H|1)=P(H|2)$ for both output ports. We find that the compensation angle is $0.0407\degree$ for the $+$ output port and $-0.2739\degree$ for the $-$ output port. The results of $P(H|\pm)$ shown in Fig.\,5 were obtained after this adjustment of the direction of the GTs by the corrsponding compensation angles was applied.

In Fig.\,5 (b), the results obtained for path 1 and for path 2 still show a slight systematic error even after the GTs have been adjusted. This difference suggests that the compensation angle is off by about $0.054\degree$ in the + port. For the evaluation of the delocalization, we used the average of the two results to obtain the value of $\theta_{0}^{2}$. Strictly speaking, the difference between the results indicates that our measurement of $P(H|+)$ is not perfectly aligned with the actual flipping direction. However, the results show that the remaining error in the observed probability $P(+|H)$ is only about $10^{-4}$, sufficiently small to be neglected in the evaluation of the data. 

\section*{Appendix B. Background Subtraction}\label{secA2}

In our experiments, we found that the effective dark counts of our detectors resulted in a constant background of about 400/s for the $+$ port and about 800/s for the $-$ port. At an intensity of about 110000/s, the Signal-to-Noise ratio (S/N) is 16.5 and 11.7 for the detectors, which correspond to 0.06 and 0.08 in precision, respectively. To achieve a precision of order $10^{-4}$, an intensity of more than $10^{10}$/s would be necessary, but such a high intensity would exceed the operating range of the single photon detectors used in the present experiment. To overcome this problem, we carefully measured the background counts and subtracted them from the counts observed in the experiment reported in this paper.

\begin{table}[h]
\renewcommand{\tablename}{Supplementary Table}
    \centering
    \caption{Background counts in the measurement time of 100s}
    \label{tab:background}
    \begin{tabular}{ccccccc}
    \hline
      & \multicolumn{3}{c}{$+$port} & \multicolumn{3}{c}{$-$port} \\ 
      & interference & Path1 & Path2 & interference & Path1 & Path2 \\ \hline
    H-direction & 44423 & 41639 & 35155 & 85951 & 36524 & 70680 \\ 
    V-direction & 48984 & 45836 & 33212 & 87425 & 42704 & 67939 \\ \hline
    \end{tabular}
\end{table}

We performed background runs after each experiment, leaving the configuration unchanged while blocking the source. Separate background runs were performed for both of the experiments where one path was blocked, and for the interference experiment where both paths were open. The measurement time was 100s, performed separately for each polarization at a phase set to $\phi=0$. Supplementary Table \ref{tab:background} shows the summary of the different background counts obtained for the different setups and the different polarization settings. The values in the table were used for background subtraction to obtain the results shown in Figs.\,4, 5, and 6.

\section*{Appendix C. Conditional delocalization}\label{secA3}

The results presented here are consistent with textbook quantum mechanics and can be predicted theoretically from an analysis of the weak interaction between a system and a quantum probe as shown in \cite{Hof21,Lem22,Hof23}. The quantity $A^2$ observed when the photon is detected in the output ports of the interferometer associated with constructive ($+$) interference or with destructive ($-$) interference at a phase shift of $\phi=0$ is determined by the small amount of entanglement between the paths and the polarization generated by the polarization rotation, as explained in \cite{Hof23}. For an input state $\mid \psi \rangle$, the value of $A^2$ conditioned by a detection described by a measurement projection $\hat{E}(f)$ is given by
\begin{equation}
A^2(f) = \frac{\langle \psi \mid \hat{A} \; \hat{E}(f) \; \hat{A} \mid \psi \rangle}{\langle \psi \mid \hat{E}(f) \mid \psi \rangle} 
\end{equation}
As was first shown in \cite{Hof21}, the operator ordering in this formula is a direct consequence of the physics describing the interaction between the system and the probe qubit. 

Textbook quantum mechanics fails to explain the statistical correlations between measurements carried out at different strengths. However, these correlations are a fundamental part of the physics described by the formalism. The focus on eigenstates as representatives of precise measurements has obscured the fact that the quantum formalism describes well-defined correlations between physical properties that cannot be observed jointly. These correlations can be observed in new kinds of experiments, and the present work is a particularly striking application of this new approach. 

In the case of an interference measurement, the outcomes $\hat{E}(+)=\mid + \rangle \langle + \mid$ and $\hat{E}(-) = \mid - \rangle \langle - \mid$ are projections of equal superpositions of the states representing the two paths. The commutation relations of $\hat{E}(f)$ and $\hat{A}$ can be given by
\begin{eqnarray}
\hat{A} \; \hat{E}(-) \; \hat{A} &=& \hat{E}(+)
\nonumber \\
\hat{A} \; \hat{E}(+) \; \hat{A} &=& \hat{E}(-).
\end{eqnarray}
Here, it is essential that operators change the states they act on. This aspect of the mathematics is actually a representation of the measurement dependence of physical reality. In the case of individual particles in an interferometer, the quantity $A^2(\pm)$ describing the delocalization of a particle detected in interference is given by the ratio of the probabilities observed in the output, with the denominator given by the probability of the observed output and the numerator given by the probability of the counterfactual alternative,
\begin{equation} 
A^2(\pm) = \frac{1-P(\pm)}{P(\pm)}.
\end{equation}
It is therefore possible to predict the experimental results based on the mathematical formalism alone. However, we would like to stress that the experiment is necessary to show that these equations describe the physical effects of particle delocalization. Hopefully, we have made it clear that it is not the goal of the present paper to ``confirm’’ a theory. Rather, we wish to demonstrate that there are observable phenomena that reveal how quantum objects behave when they are not subjected to the extreme interactions required for particle detection. Up to now, these phenomena tend to be misinterpreted because of a textbook induced bias in favor of an identification of eigenstates and eigenvalues with hypothetical realities. Hopefully, the present work will help us overcome this bias and realize the full potential of quantum physics.

\end{document}